\documentclass[twocolumn,showpacs,preprintnumbers,amsmath,amssymb,floatfix]{revtex4}

\usepackage{graphicx}
\usepackage{dcolumn}
\usepackage{bm}
\usepackage{feynmp}
\usepackage{epsfig}

\newcommand{\setval}{\fmfset{wiggly_len}{1.5mm}\fmfset{arrow_len}{1.5mm}\fmfset{arrow_ang}{13}\fmfset{dash_len}{1.5mm}\fmfpen{0.125mm}\fmfset{dot_size}{1thick}}
\newcommand{\beq}{\begin{eqnarray}}
\newcommand{\eeq}{\end{eqnarray}}
\newcommand{\hs}{\hspace{1mm}}
\newcommand{\hb}{\hbar \beta}
\newcommand{\scs}{\scriptstyle}
\newcommand{\no}{\nonumber}

\begin{document}


\title{Recursive Calculation of Effective Potential and Variational
  Resummation}

\author{Sebastian F.\ Brandt}
\email{sebastian.brandt@physik.fu-berlin.de}
\author{Hagen Kleinert}
\email{kleinert@physik.fu-berlin.de}
\affiliation{%
Freie Universit{\"a}t Berlin, Institut f{\"u}r Theoretische Physik,
Arnimallee 14, 14195 Berlin, Germany}%
\author{Axel Pelster}
\email{pelster@uni-essen.de}
\affiliation{%
Universit{\"a}t Duisburg-Essen, Campus Essen, Fachbereich Physik, Universit{\"a}tsstra{\ss}e 5, 45117 Essen, Germany}%

\date{\today}

\begin{abstract}
We set up a method for a recursive calculation of the effective potential which is
applied to a cubic potential with imaginary coupling. The
result is resummed using variational perturbation theory (VPT),
yielding an exponentially fast convergence. 
\end{abstract}


\maketitle
\section{Introduction}
\label{sec1}
Perturbation theory is the most commonly used technique for an
approximate description of non-exactly solvable systems.
However, most perturbation series are 
divergent and yield acceptable results only after resummation. In recent years, based on a variational approach due to Feynman and Kleinert \cite{Feynman2}, a
systematic and uniformly convergent {\it variational perturbation
  theory} (VPT) has been developed
\cite{Kleinertsys,PathInt3,VerenaBuch,Festschrift}. It permits the conversion of  divergent weak-coupling into convergent
strong-coupling expansions and has been applied successfully in
quantum mechanics, quantum statistics, condensed matter physics, and
the theory of critical phenomena.\\ \\  
The convergence of VPT has been proved to be exponentially fast
\cite{PathInt3,VerenaBuch}, and this has been verified for the
ground-state energy of different quantum mechanical model systems. If
the underlying potential is mirror-symmetric, one introduces a trial
oscillator whose frequency $\Omega$ is regarded as a variational parameter
and whose influence is minimized according to the {\it principle of
  minimal sensitivity} \cite{Stevenson}. In this way, the ground-state
energy of the quartic anharmonic oscillator was analyzed up to very
high orders in Refs.\ \cite{JankeC1,JankeC2}. \\ \\
If the potential is not mirror-symmetric, the center of
fluctuations no longer lies at the origin but at some nonzero place
$X$. In VPT, this situation is accounted for by regarding the
nonvanishing center of fluctuations $X$ as a second variational
parameter. An extreme example is a complete antisymmetric potential,
such as $V(x) = Ax^3$, which for real $A$ does not correspond to a
stable system. Interestingly, if the parameter $A$ is chosen to be imaginary, 
so that there does not exist a classical system at all, the
quantum-mechanical system turns out to be well-defined, and 
the spectrum of the Hamilton operator
\beq
H = - \frac{1}{2}\frac{\partial^2}{\partial x^2} + ix^3 \label{HamBen}
\eeq
is real and positive. This remarkable property of the
non-Hermitian Hamilton operator, found in Refs.\ \cite{BenderPT,BenderPT2,BenderPT3,BenderPT4,BenderPT5}, can be attributed to the fact that it
possesses a different symmetry: it is invariant under the combined
application of the parity and the time-reversal operation. 
\\
\\
In this paper, we apply VPT to the ${\cal PT}$-symmetric Hamilton
operator (\ref{HamBen}). In a first naive approach, we ignore the necessary
shift $X$ of the center of fluctuations and resum the weak-coupling series of
the ground-state energy for the anharmonic oscillator 
\beq
V(x) = \frac{\omega^2}{2} x^2 + igx^3 \label{Anhigx3}
\eeq
in the strong-coupling limit. In this limit, the potential
(\ref{Anhigx3}) reduces to the purely cubic potential of (\ref{HamBen}). 
It turns out that the VPT results approach the corresponding
numerical value for the ground-state energy of (\ref{HamBen})
with increasing order, but the
rate of convergence is not satisfactory. 
Afterwards, we allow for a nonvanishing center of fluctuations $X$ by
using the effective potential, whose calculation is accomplished by an
efficient recursion scheme. This refined approach improves the
convergence of the results drastically. 
\\
\\
In Section \ref{sec2}, we derive the weak-coupling series
for the ground-state energy of (\ref{Anhigx3}) by evaluating connected
vacuum diagrams.
In Section \ref{sec3}, we show how
this perturbation series can be obtained more efficiently by means of
the Bender-Wu recursion method \cite{Bender/Wu}. 
In Section \ref{sec4}, we resum the weak-coupling
series for the ground-state energy of (\ref{Anhigx3}) by applying VPT
and examine the resulting convergence. 
In Section \ref{sec5}, we determine the effective
potential with the {\it background method} \cite{DeWitt,Jackiw}
from one-particle irreducible vacuum diagrams. 
In Section \ref{sec6}, we set up new recursion relations for a more
efficient calculation of the effective potential.
In Section \ref{sec7}, we finally treat the resulting expansion with VPT and examine the improved convergence. 
%
%
%
%
%
%
\section{Perturbation Theory}
\label{sec2}
\begin{fmffile}{graph}
\setlength{\unitlength}{1mm}
The perturbation series
for the ground-state energy of the anharmonic oscillator (\ref{Anhigx3})
can be calculated from connected vacuum diagrams. Up to the fourth
order in the coupling constant $g$, the ground-state energy is given
by the Feynman diagrams
\beq
E & = & \frac{\hbar \omega}{2} -\lim_{T \to 0} k_BT \bigg\{ \hs
\frac{1}{8} \hspace*{2mm}
\parbox{15mm}{\centerline{
\begin{fmfgraph}(12,4)
\setval
\fmfforce{0w,0.5h}{v1}
\fmfforce{1/3w,0.5h}{v2}
\fmfforce{2/3w,0.5h}{v3}
\fmfforce{1w,0.5h}{v4}
\fmf{plain,left=1}{v1,v2,v1}
\fmf{plain,left=1}{v3,v4,v3}
\fmf{plain}{v2,v3}
\fmfdot{v2,v3}
\end{fmfgraph} } }
+ \frac{1}{12} \hspace*{2mm}
\parbox{7mm}{\centerline{
\begin{fmfgraph}(4,4)
\setval
\fmfforce{0w,0.5h}{v1}
\fmfforce{1w,0.5h}{v2}
\fmf{plain,left=1}{v1,v2,v1}
\fmf{plain}{v1,v2}
\fmfdot{v1,v2}
\end{fmfgraph} } }
\label{dia3l} \\ &&
\hspace{-3mm} + \hs \frac{1}{24}\hspace*{2mm}
\parbox{9mm}{\centerline{
\begin{fmfgraph}(4,4)
\setval
\fmfforce{0w,0h}{v1}
\fmfforce{1w,0h}{v2}
\fmfforce{1w,1h}{v3}
\fmfforce{0w,1h}{v4}
\fmf{plain,right=1}{v1,v3,v1}
\fmf{plain}{v1,v3}
\fmf{plain}{v2,v4}
\fmfdot{v1,v2,v3,v4}
\end{fmfgraph} } } 
+
\frac{1}{16}\hspace*{2mm}
\parbox{9mm}{\centerline{
\begin{fmfgraph}(4,4)
\setval
\fmfforce{0w,0h}{v1}
\fmfforce{1w,0h}{v2}
\fmfforce{1w,1h}{v3}
\fmfforce{0w,1h}{v4}
\fmf{plain,right=1}{v1,v3,v1}
\fmf{plain,right=0.4}{v1,v4}
\fmf{plain,left=0.4}{v2,v3}
\fmfdot{v1,v2,v3,v4}
\end{fmfgraph} } } 
+ \frac{1}{8}\hspace*{2mm}
\parbox{15mm}{\centerline{
\begin{fmfgraph}(12,4)
\setval
\fmfforce{0w,1/2h}{v1}
\fmfforce{1/3w,1/2h}{v2}
\fmfforce{2/3w,1/2h}{v3}
\fmfforce{5/6w,0h}{v4}
\fmfforce{5/6w,1h}{v5}
\fmfforce{1w,1/2h}{v6}
\fmf{plain,left=1}{v1,v2,v1}
\fmf{plain,left=1}{v3,v6,v3}
\fmf{plain}{v2,v3}
\fmf{plain}{v4,v5}
\fmfdot{v2,v3,v4,v5}
\end{fmfgraph} } } 
\no \\  && \hspace{-3mm}
+ \hs \frac{1}{16}\hspace*{2mm}
\parbox{23mm}{\centerline{
\begin{fmfgraph}(20,4)
\setval
\fmfforce{0w,1/2h}{v1}
\fmfforce{1/5w,1/2h}{v2}
\fmfforce{2/5w,1/2h}{v3}
\fmfforce{3/5w,1/2h}{v4}
\fmfforce{4/5w,1/2h}{v5}
\fmfforce{1w,1/2h}{v6}
\fmf{plain,left=1}{v1,v2,v1}
\fmf{plain,left=1}{v3,v4,v3}
\fmf{plain,left=1}{v5,v6,v5}
\fmf{plain}{v2,v3}
\fmf{plain}{v4,v5}
\fmfdot{v2,v3,v4,v5}
\end{fmfgraph} } } 
+\frac{1}{48}\hspace*{2mm}
\parbox{17mm}{\centerline{
\begin{fmfgraph}(13.856,12)
\setval
\fmfforce{0w,0h}{v1}
\fmfforce{1/4w,1/6h}{v2}
\fmfforce{1/2w,1/3h}{v3}
\fmfforce{3/4w,1/6h}{v4}
\fmfforce{1w,0h}{v5}
\fmfforce{1/2w,2/3h}{v6}
\fmfforce{1/2w,1h}{v7}
\fmf{plain,left=1}{v1,v2,v1}
\fmf{plain,left=1}{v4,v5,v4}
\fmf{plain,left=1}{v6,v7,v6}
\fmf{plain}{v2,v3}
\fmf{plain}{v4,v3}
\fmf{plain}{v3,v6}
\fmfdot{v2,v3,v4,v6}
\end{fmfgraph} } } 
 + {\cal O}(g^6) \bigg\} \hs, \no 
\eeq
with the propagator
\beq
  \parbox{7mm}{\centerline{
  \begin{fmfgraph*}(6,6)
  \setval
  \fmfforce{0.w,0.5h}{v1}
  \fmfforce{1w,0.5h}{v2}
  \fmf{plain}{v1,v2}
  \fmfv{decor.size=0, label=${\scs i}$, l.dist=1mm, l.angle=-180}{v1}
  \fmfv{decor.size=0, label=${\scs j}$, l.dist=1mm, l.angle=0}{v2}
  \fmfdot{v1,v2}
  \end{fmfgraph*} } } 
\hspace*{5mm}
\to  \hspace*{2mm} G_{\omega}(\tau_i, \tau_j) = \frac{\hbar}{2\omega}e^{-\omega|\tau_i - \tau_j|} \label{FeynRuleAll1}
\eeq
and the vertices
\beq
  \parbox{7mm}{\centerline{
  \begin{fmfgraph*}(6,6)
  \setval
  \fmfforce{0.5w,1h}{v1}
  \fmfforce{0.w,0.h}{v2}
  \fmfforce{1w,0.h}{v3}
  \fmfforce{1/2w,1/2h}{v4}
  \fmf{plain}{v1,v4}
  \fmf{plain}{v2,v4}
  \fmf{plain}{v3,v4}
  \fmfv{decor.size=0, label=${\scs i}$, l.dist=1mm, l.angle=-90}{v4}
  \fmfdot{v4}
  \end{fmfgraph*} } } 
\hspace*{5mm}
\to \hspace*{2mm} -\frac{6ig}{\hbar} \int_0^{\hb}d\tau_i \hs.  \label{FeynRuleAll2}
\eeq
Evaluating the Feynman diagrams (\ref{dia3l}) leads to the following analytical
expression for the ground-state energy:
\beq
E = \hbar \omega \left[ \frac{1}{2} + \frac{11g^2\hbar}{8\omega^5} -
\frac{465g^4\hbar^2}{32\omega^{10}} + {\cal O}(g^6) \right] \label{PerExpigx3} \hs.
\eeq
Since evaluating Feynman diagrams of higher orders is cumbersome, only
low perturbation orders are feasible by this procedure. If we want
to study higher orders, we better use the Bender-Wu recursion relations
\cite{Bender/Wu}.
\section{Bender-Wu Recursion Relations}
\label{sec3}
The Schr{\"o}dinger eigenvalue equation for the anharmonic oscillator
(\ref{Anhigx3}),
\beq
-\frac{\hbar^2}{2}\psi''(x) + \left(\frac{\omega^2}{2}x^2 + igx^3\right)\psi(x) =
E\psi(x)\hs, \label{Seqigx3}
\eeq
is solved as follows: We write the wave function $\psi(x)$ as 
\beq
\psi(x) =  \left(\frac{\omega}{\pi \hbar}\right)^{1/4} \exp \left[ - \frac{\hat{x}^2}{2} + \phi(\hat{x}) \right]\hs, \label{CAns}
\eeq
with the abbreviation $\hat{x} = x\sqrt{\omega/ \hbar}$,
and expand the exponent in powers of the dimensionless coupling constant $\hat{g} = g\sqrt{\hbar/ \omega^5}$ by using
\beq
\phi(\hat{x}) = \sum_{k = 1}^{\infty}\hat{g}^k \phi_k(\hat{x}) \hs.\label{PhiAns}
\eeq
The $\phi_k(\hat{x})$ are expanded in powers of the rescaled coordinate $\hat{x}$:
\beq
\phi_k(\hat{x}) = \sum_{m = 1}^{k +2}  c_m^{(k)} \hat{x}^m\hs. \label{PhikAns}
\eeq
For the ground-state energy we make the ansatz
\beq
E = \hbar\omega \left( \frac{1}{2} + \sum_{k = 1}^{\infty} \hat{g}^k \epsilon_k \right) \hs. \label{EnAns}
\eeq
Inserting (\ref{CAns}) -- (\ref{EnAns}) into (\ref{Seqigx3}),
we obtain to first order
\beq
c_1^{(1)} = - i, \quad c_2^{(1)} = 0, \quad  c_3^{(1)} = -
\frac{i}{3}, \quad  \epsilon_1 = 0 \hs. \label{eps1}
\eeq
For $k \geq 2$, we find the following recursion relation for the
expansion coefficients in (\ref{PhikAns}): 
\beq
c_m^{(k)} &=& \frac{(m+2)(m+1)}{2m}\hs c_{m + 2}^{(k)} 
\label{cmk} \\ &&
+ \hs \frac{1}{2m}\sum_{l = 1}^{k-1}\sum_{n =
  1}^{m+1} n(m+2-n) \hs c_n^{(l)}\hs c_{m +2 - n}^{(k - l)}\hs, \no
\eeq
with $c_m^{(k)} \equiv 0$ for $m > k + 2$. The expansion coefficients of
the ground-state energy follow from
\beq
\epsilon_k =  - c_2^{(k)} -\frac{1}{2}\sum_{l =
  1}^{k-1}c_1^{(l)}\hs c_1^{(k-l)}\hs. \label{ek}
\eeq
Table \ref{EpsTable} shows the coefficients $\epsilon_k$ up to the
10th order. 
We observe that all odd orders vanish for symmetry reasons and that
the first two even orders agree, indeed, with the earlier result
(\ref{PerExpigx3}).
\begin{table}[t]
\begin{center}
\begin{tabular}{|c | c c c c c|}
\hline
\rule[-1mm]{0mm}{5mm} $k$  & $1$ & $2$ & $3$ & $4$ & $5$ \\[1mm]
\hline
\rule[-1mm]{0mm}{7mm} $\epsilon_k$ & $0$ &  ${\displaystyle \frac{11}{8}}$ & $0$ &
${\displaystyle -\frac{465}{32} }$ & $0$ \\[3mm] \hline \hline
\rule[-1mm]{0mm}{5mm}$k$ & $6$ & $7$ & $8$ & $9$ & $10$ \\[1mm]
\hline
\rule[-1mm]{0mm}{7mm}
$\epsilon_k$ & \hspace{1mm} ${\displaystyle  \frac{39709}{128} }$
\hspace{1mm} & $0$ & $
{\displaystyle - \frac{19250805}{2048} }$ & $0$ &
${\displaystyle \frac{2944491879}{8192} } \hspace{2mm} $ \\[3mm] \hline
\end{tabular}
\caption{Expansion coefficients for the ground-state energy of the anharmonic oscillator
  (\ref{Anhigx3}) up to the 10th order.}\label{EpsTable}
\end{center}
\end{table}
\section{Resummation of Ground-State Energy}
\label{sec4}
In this section, we consider the strong-coupling limit of the
perturbation series (\ref{EnAns}). Rescaling the
coordinate according to $x \to xg^{-1/5}$, 
the Schr{\"o}dinger equation (\ref{Seqigx3}) becomes
\beq
-\frac{\hbar^2}{2}\hs\frac{\partial^2}{\partial x^2} \hs \psi(x) + \left( \frac{1}{2}g^{-4/5}\omega^2x^2 + ix^3
\right)\psi(x) &=& \label{SchrEqReSca} \\
&&\hspace{-10mm} g^{-2/5}E \hs \psi(x)\hs. \no
\eeq
Expanding the wave function and the energy in powers of the coupling
constant yields
\beq
\psi(x) = \psi_0(\hat{x}) + \hat{g}^{-4/5}\psi_1(\hat{x}) + \hat{g}^{-8/5}\psi_2(\hat{x}) + \ldots
\eeq
and
\beq
E &=&  \hbar \omega  \hs \hat{g}^{2/5}  \label{StrCouReSca} \left(b_0 +  \hat{g}^{-4/5}b_1 + \hat{g}^{-8/5}b_2 + \ldots
\right) \hs.  
\eeq
By considering (\ref{SchrEqReSca}) in the limit $g\to \infty$,
we find that the leading strong-coupling coefficient $b_0$ equals
the ground-state energy associated with the  Hamilton operator (\ref{HamBen}).
A precise numerical value for this ground-state energy was given by
C.M. Bender \cite{BenderPT,BenderPrivat}:
\beq
b_0 =  0.762851773\ldots \hs. \label{Eexa}
\eeq
The weak-coupling series (\ref{EnAns}) is of the form
\beq
E^{(N)}(\alpha, \omega) = \hbar \omega \left[ \frac{1}{2} +  \sum_{k = 1}^N \left(\frac{\hbar \alpha}{\omega^5}\right)^k\epsilon_{2k}\right]  \hs, \label{Eexp}
\eeq
with the abbreviation $\alpha = g^2$. Table \ref{EpsTable} suggests
that (\ref{Eexp}) represents a divergent Borel series which is
resummable 
by applying VPT
\cite{Kleinertsys,PathInt3,VerenaBuch,Festschrift}. 
To this end, an artificial parameter is introduced in the perturbation
series, which is most easily obtained by Kleinert's square-root trick
\beq
\omega \to \Omega \sqrt{1 + \alpha  r}\hs, \label{KleTrick1}
\eeq
with
\beq
r = \frac{\omega^2 - \Omega^2}{ \alpha \Omega^2}\hs. \label{KleTrick2}
\eeq
Thus, one replaces the frequency $\omega$ in the weak-coupling series
(\ref{Eexp}) according to (\ref{KleTrick1})
and re-expands the resulting expression in powers of $\alpha$ up to the order
$\alpha^{N}$. Afterwards,
the auxiliary parameter $r$ is replaced according to
(\ref{KleTrick2}). The ground-state energy thus becomes dependent on
the variational parameter $\Omega$: $E^{(N)}(\alpha,\hs \omega) \to
E^{(N)}(\alpha,\hs \omega, \hs \Omega)$. The
influence of $\Omega$ is then optimized according to the principle of minimal
  sensitivity \cite{Stevenson}, i.e.\ one approximates the
  ground-state energy to $N$th order by
\beq
E^{(N)} = E^{(N)}(\alpha, \hs \omega, \hs \Omega^{(N)})\hs,
\eeq
where $\Omega^{(N)}$ denotes that value of the variational parameter for
which $E^{(N)}(\alpha, \hs \omega, \hs \Omega)$ has an extremum or a turning point.
\\ \\
Consider, as an example, the weak-coupling series (\ref{Eexp}) to first order:
\beq
E^{(1)}(\alpha, \hs \omega ) = \frac{\hbar\omega}{2} + \alpha \frac{11\hbar^2}{8\omega^4}\hs.  
\eeq
Inserting
(\ref{KleTrick1}), re-expanding in $\alpha$ to first order, and
taking into account (\ref{KleTrick2}), we obtain
\beq
E^{(1)}(\alpha, \hs \omega, \hs \Omega) = \frac{\hbar\Omega}{4} + \frac{\hbar\omega^2}{4\Omega} +
\alpha \frac{11\hbar^2}{8\Omega^4}\hs . \label{E1Om}
\eeq
Extremizing this and going to large coupling constants, we obtain the strong-coupling behavior of the variational
parameter: 
\beq
\hspace{-2mm} \Omega^{(1)} = \omega \hat{\alpha}^{1/5}\left(\Omega^{(1)}_0 + \Omega^{(1)}_1 \hat{\alpha}^{-2/5} +
  \Omega^{(1)}_2 \hat{\alpha}^{-4/5} + \ldots \right), \label{StrCouOm}
\eeq  
with the abbreviation $\hat{\alpha} = \hat{g}^2$ and the coefficients 
\beq
\hspace{-1.5mm}\Omega_0^{(1)} = \sqrt[5]{22},  \hs  \Omega_1^{(1)} =
\frac{1}{5 \sqrt[5]{22}},  \hs  \Omega_2^{(1)} =
\frac{1}{25\sqrt[5]{10648}}, \hs \ldots
\hs.   \label{OmCoeff}
\eeq
Inserting the result (\ref{StrCouOm}), (\ref{OmCoeff}) into
(\ref{E1Om}), we obtain the strong-coupling
series (\ref{StrCouReSca}) with the first-order coefficients
\beq
\hspace{-1mm} b_0^{(1)} = \frac{5\sqrt[5]{22}}{16}, \hs \hs
b_1^{(1)} = \frac{4}{\sqrt[5]{22}}, \hs \hs
b_2^{(1)} = \frac{-1}{100 \sqrt[5]{22}}
\hs, \hs \ldots \hs.  \label{StrCouCoex3}
\eeq
The numerical value of the leading strong-coupling coefficient is $b_0^{(1)}
\approx 0.5799$. Thus, to first order, the relative deviation of the
result from the precise value (\ref{Eexa}) is
\beq
\frac{|b^{(1)}_0 - b_{0}|}{b_{0}} \approx 24 \% \hs. \label{x3badresult}
\eeq
Despite this relatively poor agreement, it turns out that the
VPT results for $b_0^{(N)}$ in higher orders converge towards the exact value
(\ref{Eexa}). In Refs.\ \cite{PathInt3,VerenaBuch} it is proved
that VPT in general yields approximations whose relative deviation from the
exact value vanishes exponentially. In our case we have
\beq
\frac{|b_0^{(N)} - b_0|}{b_0} \propto \exp\left(-C N^{3/5} \right)\hs, \label{KonVer}
\eeq
where the exponent $3/5$ is determined by the structure of the
strong-coupling series (\ref{StrCouReSca}). 
\\ \\
In Fig.\ \ref{figcomp} the exponential convergence of our variational results is shown up to the $20$th
order. Fitting the logarithm of the relative deviation to a straight
line yields
\beq
\ln \frac{|b_0^{(N)} - b_0|}{b_0} = -0.96(11)N^{3/5} -
1.83(44)\hs.  \label{FitReswoVeff}
\eeq
In the following, we show how this exponential convergence is
improved drastically by allowing for a shift of the center of fluctuations.
\section{Diagrammatic Approach to Effective Potential}\label{sec5}
In the presence of a constant external current $j$, the quantum
statistical partition function reads
 \beq
Z(j) := \oint{\cal D}x\hs\exp\left\{-\frac{1}{\hbar}\hs{\cal A}[x](j) 
\right\} \hspace{0.5mm},\label{ZjC}
\eeq
where ${\cal A}[x](j)$ is the Euclidean action:
\beq
{\cal A}[x](j) = \int_{0}^{\hbar\beta}d\tau\left[\frac{1}{2}\dot{x}^{2}(\tau) +
  V(x(\tau)) -jx(\tau) \right] \hs \label{ImTimAct}.
\eeq
The free energy thus becomes a function of the external current:
\beq
F(j) = -\frac{1}{\beta}\ln Z(j)\hs.
\eeq
The path average,
\beq
X =\frac{1}{Z(j)} \oint{\cal D}x  \left[
  \int_0^{\hb}\frac{d\tau}{\hb}\hs x(\tau)\hs\right] \exp\left\{-\frac{1}{\hbar}\hs{\cal A}[x](j)
\right\} \hs, \hspace{-2mm} \no \\ 
\eeq
then follows from the first derivative of the free
energy with respect to the external current:
\beq
X = -\frac{\partial F(j)}{\partial j}\hs. \label{Xder}
\eeq
Assuming that the last identity can be inverted to yield the current
$j$ as a function of the average $X$, one defines the effective
potential $V_{\rm eff}(X)$ as the Legendre transform of the free
energy with respect to the external current:
\beq
V_{\rm eff}(X) = F(j(X)) + j(X)X\hs. \label{EffPot}
\eeq 
Furthermore, the first derivative of the effective potential gives back the external current $j$:
\beq
\frac{\partial V_{\rm eff}(X)}{\partial X} = 
j(X)\hs. \label{CurDerX}
\eeq 
Thus, the free energy $F \equiv F(j = 0)$ can be obtained by extremizing
the effective potential,
\beq
F = V_{\rm eff}(X_{\rm e}) \hs,
\eeq
with
\beq
\left. \frac{\partial V_{\rm eff}(X)}{\partial X} \right|_{X = X_{\rm e}} = 0\hs.
\label{VeffExtCond}
\eeq
In the zero-temperature limit, extremizing the effective potential then
yields the ground-state energy. \\ \\
The effective potential is usually not calculated by
performing explicitly the Legendre transformation (\ref{EffPot}) but
by a diagrammatic technique derived via the so-called background
method \cite{Jackiw,DeWitt}.
There, the effective potential is expanded in powers of the Planck
constant $\hbar$, and the expansion terms are one-particle irreducible
vacuum diagrams. The result is
\beq
V_{\rm eff}(X) = V(X) + \frac{\hbar}{2} \mbox{Tr} \ln G^{-1} + V^{(\rm
  int)}(X) \hs, \label{VeffBacMet}
\eeq
where the trace-log term is given by the ground-state energy of a
harmonic oscillator of $X$-dependent frequency $\tilde{\omega} = \sqrt{V''(X)}$:
\beq
\frac{\hbar}{2} \mbox{Tr} \ln G^{-1} = \frac{\hbar \tilde{\omega}}{2} \hs.
\eeq
The interaction term $V^{(\rm int)}(X)$ contains the sum of all
one-particle irreducible vacuum diagrams. For the anharmonic
oscillator (\ref{Anhigx3}), the relevant subset of the diagrams in
(\ref{dia3l}) is
\beq
V^{(\rm int)}(X) &=& - \lim_{T \to 0} k_BT
\bigg\{ \hs
\frac{1}{12} \hspace*{2mm}
\parbox{7mm}{\centerline{
\begin{fmfgraph}(4,4)
\setval
\fmfforce{0w,0.5h}{v1}
\fmfforce{1w,0.5h}{v2}
\fmf{plain,left=1}{v1,v2,v1}
\fmf{plain}{v1,v2}
\fmfdot{v1,v2}
\end{fmfgraph} } }
+ \frac{1}{24}\hspace*{2mm}
\parbox{9mm}{\centerline{
\begin{fmfgraph}(4,4)
\setval
\fmfforce{0w,0h}{v1}
\fmfforce{1w,0h}{v2}
\fmfforce{1w,1h}{v3}
\fmfforce{0w,1h}{v4}
\fmf{plain,right=1}{v1,v3,v1}
\fmf{plain}{v1,v3}
\fmf{plain}{v2,v4}
\fmfdot{v1,v2,v3,v4}
\end{fmfgraph} } } 
\no \\
&& 
+ \hs
\frac{1}{16}\hspace*{2mm}
\parbox{9mm}{\centerline{
\begin{fmfgraph}(4,4)
\setval
\fmfforce{0w,0h}{v1}
\fmfforce{1w,0h}{v2}
\fmfforce{1w,1h}{v3}
\fmfforce{0w,1h}{v4}
\fmf{plain,right=1}{v1,v3,v1}
\fmf{plain,right=0.4}{v1,v4}
\fmf{plain,left=0.4}{v2,v3}
\fmfdot{v1,v2,v3,v4}
\end{fmfgraph} } } 
\hspace{-1mm} \bigg\} + \hs {\cal O}(\hbar^4) \label{Veff3l} \hs.
\eeq
These one-particle irreducible vacuum diagrams are derived most easily
by an efficient graphical recursion method \cite{Phi3Phi4}.
The frequency of the propagators is now given by:
\beq 
\tilde{\omega} = \sqrt{\omega^2 + 6igX} \hs.
\eeq
By evaluating the diagrams (\ref{Veff3l}) we obtain
\beq 
V_{\rm eff}(X) &=& \frac{\omega^2}{2}X^2 + igX^3 + \frac{\hbar}{2}
\sqrt{\omega^2 + 6igX}  \label{Veff3h} \\ && \hspace{-10mm} + \frac{\hbar^2g^2}{4(\omega^2 + 6igX )^2} -
\frac{51 \hbar^3 g^4}{ 32 \left( \omega^2 + 6igX
    \right)^{9/2} }  +
{\cal O}(\hbar^4) \hs. \no 
\eeq
The ground-state energy of the anharmonic oscillator (\ref{Anhigx3})
is found by extremizing the effective potential (\ref{Veff3h}). 
To this end, we expand the extremal background according to
\beq
X_{\rm e} = i(X_0 + \hbar X_1 + \hbar^2 X_2 +\hbar^3X_3) + {\cal O}( \hbar^4 ) \hs. \label{XextAns}
\eeq
Inserting (\ref{XextAns}) into the vanishing first derivative of
(\ref{Veff3h}) and re-expanding in $\hbar$, we obtain a system of equations which are solved by
\beq
X_0 = 0\hs, \quad \quad X_1 = -\frac{3g}{2\omega^3}\hs, \quad \quad X_2 =
\frac{33g^3}{2\omega^8}\hs. \label{X0X1X2}
\eeq
Inserting (\ref{XextAns}), (\ref{X0X1X2}) into (\ref{Veff3h}) and
re-expanding in $\hbar$ yields again the ground-state energy
(\ref{PerExpigx3}).
\\
\\
In order to go to higher orders, we shall now
develop a recursion relation for the effective potential.
\section{Recursive Approach to Effective Potential} \label{sec6}
In the presence of a constant external current $j$, the Schr{\"o}dinger eigenvalue
equation for the anharmonic oscillator (\ref{Anhigx3}) reads
\beq
\hspace{-1mm} -\frac{\hbar^2}{2}\psi''(x) + \left(\frac{\omega^2}{2}x^2 + igx^3 -jx\right)\psi(x) =
E\psi(x) \hs.\label{Seqigx3j}
\eeq
Taking into account the Legendre identities (\ref{EffPot}), (\ref{CurDerX}),
Eq.\ (\ref{Seqigx3j}) becomes
\beq
-\frac{\hbar^2}{2}\psi''(x) + \left(\frac{\omega^2}{2}x^2 + igx^3 -V_{\rm
    eff}'(X)x\right)\psi(x) \no 
\\ = \Big[V_{\rm eff}(X) - V_{\rm eff}'(X)X\Big]\psi(x)\hs. \label{Seqigx3Veff}
\eeq 
If the coupling constant $g$ vanishes, Eq.\ (\ref{Seqigx3Veff}) is solved by 
\beq
\psi(x) &=& {\cal N}\hs \exp \left(\hat{X}\hat{x} - \frac{\hat{x}^2}{2} \right)\hs,\\
V_{\rm eff}(X) &=& \hbar\omega \left( \frac{1}{2} + \frac{\hat{X}}{2}
\right) \hs,
\eeq
where the path average has been rescaled by the oscillator length:
$\hat{X} = X\sqrt{\omega / \hbar}$.
For a nonvanishing coupling constant $g$, we solve the
differential equation (\ref{Seqigx3Veff}) by 
the expansions
\beq
\psi(x) &=& {\cal N}\hs \exp \left[\hat{X}\hat{x} - \frac{\hat{x}^2}{2} +
  \phi(\hat{x})\right],  \label{PsiAnsWcX}\\ 
V_{\rm eff}(X) &=&   \hbar\omega \left[ \frac{1}{2} + \frac{\hat{X}^2}{2} +
\sum_{k = 1}^{\infty} \hat{g}^k V_k(\hat{X}) \right] \hs. \label{EnExpWcX}
\eeq
For the correction to the wave function, $\phi(\hat{x})$, we make again the
ansatz (\ref{PhiAns}), (\ref{PhikAns}). 
Thus, we obtain from (\ref{Seqigx3Veff})
for $k = 1$:
\beq
&& \hspace{-9mm}c_1^{(1)} =  \frac{i}{2} + 2i\hat{X}^2, \quad
\hs \hs
c_2^{(1)} = -\frac{i\hat{X}}{2}, \quad \hs \hs c_3^{(1)} = -
\frac{i}{3}\hs, \label{c11EffPot} \\   
&& \hspace{14mm} V_1(\hat{X}) =  \frac{3 i \hat{X}}{2}
+i \hat{X}^3 \hs. \label{V1EffPot}
\eeq
For $k \geq 2$ one finds for $m \geq 2$ the following recursion relation for
the expansion coefficients of the wave function
\beq
c_m^{(k)} &=& \frac{(m+2)(m+1)}{2m}\hs c_{m + 2}^{(k)}  +
\frac{\hat{X}(m+1)}{m} \hs c_{m+1}^{(k)} \no  \\ && \hspace{-5mm} + \hs
\frac{1}{2m}\sum_{l = 1}^{k-1}\sum_{n = 1}^{m+1} n(m+2-n) \hs c_n^{(l)}\hs c_{m +2
  - n}^{(k - l)}\hs,   \label{cmkEffPot} 
\eeq
with $c_m^{(k)} \equiv 0$ for $m > k + 2$.
For $m =1$, we have
\beq
c_1^{(k)} &=& 3 c_3^{(k)} + 2\hat{X}c_2^{(k)}  +
V_k'(\hat{X}) \no \\
&& + \sum_{l=1}^{ k - 1} \left( c_2^{(k - l)}c_1^{(l)} +
  c_1^{(k -l)}c_2^{(l)} \right)\hs. \label{c1kEffPot}
\eeq
The expansion coefficients of the effective potential follow from
\beq
V_k(\hat{X}) & = & - c_2^{(k)} -  3\hat{X} c_3^{(k)} - 2\hat{X}^2
c_2^{(k)}  \label{VkEffPot}  \\ && \hspace{-16mm} - \hat{X} \sum_{l = 1}^{k -1} \left( c_2^{(k -l)}c_1^{(l)} + c_1^{(k -l)}c_2^{(l)} \right) - \frac{1}{2}\sum_{l=1}^{k-1}c_1^{(l)}\hs c_1^{(k-l)}. \no
\eeq
\begin{table}[t]
\begin{center}
\begin{tabular}{|c | c c c| }
\hline
\rule[-1mm]{0mm}{5mm} $l$  & $0$ & $1$ & $2$  \\[1mm] \hline
\rule[-1mm]{0mm}{7mm} $V^{(l)}(X)$ &\hspace{1mm} ${\displaystyle \frac{\omega^2}{2}X^2 +
  igX^3} $   &  $ {\displaystyle 
  \frac{\tilde{\omega}}{2} }$  &  ${\displaystyle \frac{g^2}{4\tilde{\omega}^4}}
$  \\[3mm] \hline \hline
\rule[-1mm]{0mm}{5mm}$l$ & $3$ & $4$ & $5$  \\[1mm] \hline
\rule[-1mm]{0mm}{7mm}$V^{(l)}(X)$ & \hspace{1mm} ${\displaystyle -\frac{51 g^4}{32
    \tilde{\omega}^{9}} }$ 
\hspace{1mm} & $ { \displaystyle \frac{3331 g^6}{128\tilde{\omega}^{14}}}$  &$ {\displaystyle -\frac{1371477 g^8}{2048
   \tilde{\omega}^{19}} }$ \hspace{1mm} \\[3mm] \hline
\end{tabular}
\caption{Expansion coefficients for the effective potential of
  (\ref{Anhigx3}) up to five loops.}\label{VeffTabh}
\end{center}
\end{table}\\
Using these results, the effective potential can be determined
recursively, yielding an expansion in the coupling constant
$g$:
\beq
V_{\rm eff}(X) &=& \frac{\hbar\omega}{2} + \frac{\omega^2}{2}X^2 +
ig\left(\frac{3\hbar X}{2 \omega} + X^3\right) \label{Veffg4} \\ 
&& \hspace{-10mm} +\hs g^2\frac{\hbar(\hbar  + 9\omega X^2)}{4\omega^4} -
ig^3\frac{3\hbar X(4 \hbar + 9\omega X^2)}{4\omega^6} 
\no \\
&& \hspace{-10mm}  - \hs g^4\frac{3\hbar (17\hbar^2 + 288 \hbar X^2 +    270 X^4 \omega^2)}{32 \omega^9} 
+ {\cal O}(g^5)\hs. \no 
\eeq
This result is in agreement with the expansion of (\ref{Veff3h}) in
powers of $g$ and can be carried to higher orders without effort. The expansion coefficients for the $\hbar$-expansion 
\beq
V_{\rm eff}(X) = \sum_{l = 0}^N \hbar^l V^{(l)}(X) + {\cal O}(\hbar^{N+1}) \label{hExp}
\eeq
can then be obtained easily \cite{Diplomarbeit}. Iterating the
recursion relations (\ref{cmkEffPot}) -- (\ref{VkEffPot}) up to the order $g^8$, we obtain
the effective potential up to five loops as shown in Tab.\
\ref{VeffTabh}.
\section{Resummation of Effective Potential} \label{sec7}
We now apply VPT to the loop expansion of the effective potential
(\ref{hExp}). Since the Planck constant $\hbar$ is now the expansion
parameter rather than the coupling constant $g$, Kleinert's
square-root trick will be modified accordingly:
\beq
\omega \to \Omega \sqrt{1 + \hbar r}\hs, \label{KleTrickh1}
\eeq
with
\beq
r = \frac{\omega^2 - \Omega^2}{\hbar \Omega^2}\hs. \label{KleTrickh2}
\eeq
As an example, we consider again the first order:
\beq
V_{\rm eff}^{(1)}(X) =  \frac{\omega^2}{2}X^2 + igX^3   + \frac{\hbar}{2}
\sqrt{\omega^2 + 6igX}\hs.
\eeq
After substituting $\omega$ according to (\ref{KleTrickh1}),
re-expanding in $\hbar$, and taking into account (\ref{KleTrickh2}), we obtain
\beq
V_{\rm eff}^{(1)}(X, \Omega) =  \frac{\omega^2}{2}X^2 + igX^3   + \frac{\hbar}{2}
\sqrt{\Omega^2 + 6igX}\hs. \label{Vef1}
\eeq
In order to calculate an approximation for the ground-state energy, we
now optimize in $\Omega$ and extremize in $X$, yielding
\beq
\frac{\partial }{\partial \Omega} V_{\rm eff}^{(1)}(X, \Omega)\bigg|_{X = X^{(1)},\hs \Omega =
  \Omega^{(1)}} &=& 0\hs, \label{ConVefOm} \\
\frac{\partial }{\partial X} V_{\rm eff}^{(1)}(X, \Omega)\bigg|_{X = X^{(1)}, \hs \Omega =
  \Omega^{(1)}} & =& 0 \hs. \label{ConVefX}
\eeq
\begin{table}[t]
\begin{center}
\begin{tabular}{|l|l|}
\hline
\rule[-1mm]{0mm}{5mm} \hspace{1mm} 1-loop VPT \hspace{2mm} & \hspace{2mm} 0.742751023 \hspace{1mm} \\[1mm] \hline
\rule[-1mm]{0mm}{5mm} \hspace{1mm} 2-loop VPT \hspace{2mm} & \hspace{2mm} 0.764570478 \hspace{1mm}  \\[1mm] \hline
\rule[-1mm]{0mm}{5mm} \hspace{1mm} 3-loop VPT \hspace{2mm} & \hspace{2mm} 0.758783545  \\[1mm] \hline
\rule[-1mm]{0mm}{5mm} \hspace{1mm} 4-loop VPT \hspace{2mm} &
\hspace{2mm} 0.762843684 \hspace{1mm}  \\[1mm] \hline
\rule[-1mm]{0mm}{5mm} \hspace{1mm} 5-loop VPT \hspace{2mm} &
\hspace{2mm} 0.762849959 \hspace{1mm} \\[1mm] \hline \hline
\rule[-1mm]{0mm}{5mm} \hspace{1mm} Numerical  & \hspace{2mm}
0.762851773 \hspace{2mm} \\[1mm] \hline  
\end{tabular}
\caption[Variational results for the ground-state energy associated
with ${\cal H} = p^2 + ix^3$]{Variational results for the ground-state energy of
  (\ref{HamBen}) compared to the numerical result (\ref{Eexa}) of
  Refs.\ \cite{BenderPT,BenderPrivat}.} \label{Results}
\end{center}
\end{table}\\
Equation (\ref{ConVefOm}) is solved by 
\beq
\Omega^{(1)} = 0 \hs. \label{VefOm1}
\eeq
Afterwards, we obtain from (\ref{ConVefX}):
\beq
X^{(1)} + \frac{\omega^2}{3ig} + \frac{ \hbar }{2 \sqrt{6ig} (X^{(1)})^{3/2}} =
0 \hs. \label{VefX1}
\eeq
This equation allows us to determine the strong-coupling behavior of
$X$:
\beq
X^{(1)}& =& -i \hat{g}^{-1/5} \sqrt{\frac{\hbar}{\omega}}  \label{X1} \\ && \times \hs  \left( X_0^{(1)}
  + X_1^{(1)}\hat{g}^{-4/5} +X_2^{(1)} \hat{g}^{-8/5} + \ldots \right) \no \hs,
\eeq
where the coefficients read
\beq
X_0^{(1)} = \frac{1}{\sqrt[5]{24}},  \hs  X_1^{(1)} = -
\frac{2}{15}, \hs  X_2^{(1)} = \frac{\sqrt[5]{24}}{75},\hs 
 \ldots \hs. \label{X1Coe}
\eeq
Inserting the results (\ref{VefOm1}), (\ref{X1}), (\ref{X1Coe}) into
(\ref{Vef1}) yields the strong-coupling behavior of the ground-state energy
(\ref{StrCouReSca}), with the new coefficients:
\beq
\hspace{-2mm}
b_0^{(1)} = \frac{5}{2 \sqrt[5]{432}},   \hs
b_1^{(1)} = -\frac{1}{4 \sqrt[5]{18}},   \hs
b_2^{(1)} = \frac{1}{15\sqrt[5]{24}},   \hs \ldots \hs.
\eeq
The new numerical value of the leading strong-coupling coefficient is
$b_0^{(1)} \approx 0.7428$, which is in much better agreement with
(\ref{Eexa}) than the previous value of (\ref{StrCouCoex3}):
\beq
\frac{|b^{(1)}_0 - b_0|}{b_0} \approx 3 \% \hs.
\eeq
Thus, the variational path average has led to a significant
improvement of the first-order result. 
Table \ref{Results} summarizes our results for $b_0^{(N)}$
up to the fifth order.
\begin{figure}[t!]
\centerline{\includegraphics[width=8.7cm]{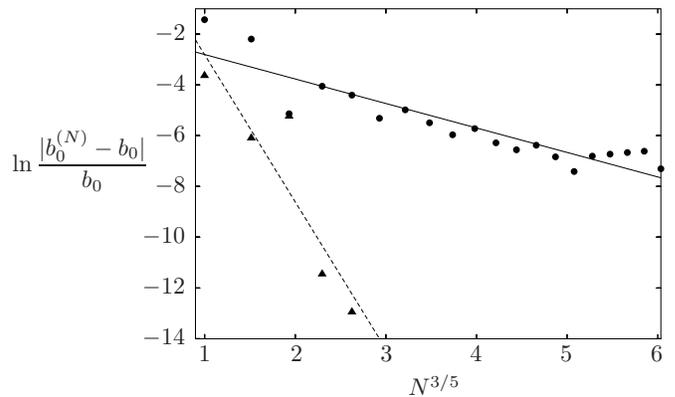}}

\caption{ \label{figcomp} Convergence of the results for the
  strong-coupling coefficient $b_0$ obtained from resummation of the
  weak-coupling series of the ground-state energy (circles). The
  resummation involving a variational path average $X$ converges
  much faster (triangles). The lines represent fits of
        the respective data to straight lines.}   
\end{figure}
Figure \ref{figcomp} shows the much faster exponential convergence up
to the fifth order.  A best fit of the data yields
\beq
\ln \frac{|b_0^{(N)} - b_0|}{b_0} = -5.8(1.6) N^{3/5} +
3.0(3.0) \hs \label{FitReswVeff}
\eeq
which is to be compared with (\ref{FitReswoVeff}).
\section{Conclusion and Outlook} \label{sec8}
We have developed a recursive technique to determine the effective
potential, which is far more efficient than diagrammatic methods. In
combination with VPT, this leads to a fast converging determination of
the ground-state energy of quantum-mechanical systems with non-mirror
symmetric potentials. It will be interesting to analyze in a similar
way systems with a coordinate dependent mass term, where only a lowest
order effective potential has been calculated so far
\cite{KleinertEffMass}. Interesting future applications will address
the effective potential of $\phi^4$ theories in $4 -\epsilon$ dimensions to
obtain equations of state near to a critical point. A first attempt in
this direction is Ref.\ \cite{Brazil}.
\begin{acknowledgments}
The authors thank C.M.\ Bender for fruitful discussions. SFB thanks
R.\ Graham for hospitality during a stay at the Universit{\"a}t
Duisburg-Essen in April 2004.
\end{acknowledgments}

%
%
\end{fmffile}
\end{document}